\documentclass[12pt]{article}

\newcommand{\pagenumber}{\pagestyle{plain}\setcounter{page}{1}} %remove for PL
\usepackage{amsmath}
\usepackage{amsfonts}
\usepackage[dvipdfm]{graphicx}
\usepackage{mathrsfs}
\usepackage{amssymb}
\usepackage{type1cm}
\usepackage{amsthm}
\theoremstyle{plain}

\theoremstyle{definition}
\newtheorem{definition}{Definition}[section]

\long\def\symbolfootnote[#1]#2{\begingroup%
\def\thefootnote{\fnsymbol{footnote}}\footnote[#1]{#2}\endgroup}

\begin{document}

\pagestyle{empty}

\begin{center}
{\large \bf Deformed Toric Ideal Constraints for Stoichiometric Networks}

\vspace{16pt}
Masamichi Sato$^{1}
\symbolfootnote[1]{Corresponding author. E-mail address: mmsato11@bi.s.u-tokyo.ac.jp}$ and Kenji Fukumizu$^{2}$% and Masanori Arita$^{1,3}$

\vspace{16pt}

{\sl $^1$Department of Biophysics and Biochemistry,\\
Graduate School of Science,\\
The University of Tokyo\\
2-11-16 Yayoi, Bunkyo-ku, Tokyo 113-0032, Japan\\
$^2$The Institute of Statistical Mathematics\\
10-3 Midori-cho, Tachikawa, Tokyo 190-8562, Japan\\
%$^3$RIKEN, Plant Science Center,\\
%1-7-22 Suehiro-cho, Tsurumi-ku, Yokohama 230-0045, Japan
}

\vspace{12pt}
{\bf ABSTRACT}

\end{center}

\begin{minipage}{4.8in}
{
We discuss chemical reaction networks and metabolic pathways based on stoichiometric network analysis, and introduce deformed toric ideal constraints by the algebraic geometrical approach.
This paper concerns steady state flux of chemical reaction networks and metabolic pathways.
With the deformed toric ideal constraints, the linear combination parameters of extreme pathways are automatically constrained without introducing ad hoc constraints. To illustrate the effectiveness of such constraints, we discuss two examples of chemical reaction network and metabolic pathway; in the former the flux and the concentrations are constrained completely by deformed toric ideal constraints, and in the latter, it is shown the deformed toric ideal constrains the linear combination parameters of flux at least partially.
Even in the latter case, the flux and the concentrations are constrained completely with the additional constraint that the total amount of enzyme is constant.

\vspace{12pt}
}
\end{minipage}

\vfill
\pagebreak

\pagestyle{plain}
\setcounter{page}{1}

\pagenumber                                     %remove for PL
\baselineskip=16pt

%%%%%%%%%%%
\section{Introduction}
%%%%%%%%%%%

Metabolic pathway analysis is one of the major fields in systems biology and is the basis of metabolic engineering and synthetic biology.
With the development of experimental technology, this field is studied with practical objectives, such as clarifying the metabolic systems of life with biological interests and manufacturing biochemical products through metabolic processes.

In chemistry, it is important to identify the chemical mechanism.
The instability of steady states of chemical reaction systems exhibits exotic dynamics, such as switching between multiple steady states, explosions and sustained oscillations.
The chemical reaction networks have been studied with the purpose of clarifying the chemical mechanism behind such phenomena.

Theoretical studies on chemical reaction networks are based on stoichiometric network analysis (SNA), which is the method based on the mass action kinetics.
SNA was initiated by Clarke \cite{clarke1,clarke2,clarke3} and succeeded by Feinberg et al.~\cite{feinberg1,feinberg2,feinberg3,feinberg4}.  Gatermann et al.~\cite{gatermann1,gatermann2,gatermann3} have studied chemical reaction networks from the viewpoint of algebraic geometry, especially of polynomial rings.
Inspired by their study, Shiu et al.~have used toric varieties in the analysis \cite{shiu1,shiu2,shiu3}.
There are also studies on bifurcations of dynamical systems with SNA~\cite{gatermann4,sensse,domijan1,domijan2}.
The relations among monomial entries of the flux vector in the context of constraining the steady-state flux space was done for MAPK cascade~\cite{conradi}.

Metabolic pathways, on the other hand, have been studied by flux balance analysis (FBA) which introduces the steady state flux space.
Our main concern is steady state flux.
The introduction of `elementary mode'~\cite{schuster1,schuster2,shilling1,schuster3,schuster4} and `extreme pathway'~\cite{palsson1} is specific progresses in this field.
Review articles on these theme include refs.~\cite{fluxreview1,fluxreview2}.
For the discussions of these two approaches, see ref.~\cite{klamt1}.

This paper discusses algebraic geometrical constraints for stoichiometric networks:
from the monomial vector expression of flux, we derive the deformed toric ideal, that works as constraints on the linear combination parameters of the flux.
The existence of such constraints had not been pointed out before.
We show two examples of analysis by the proposed method in a chemical reaction network and a metabolic pathway. In the former example, we show that all of linear combination parameters and concentrations are determined completely by the reaction coefficients with the deformed toric ideal constraints.
In the latter, we show that the linear combination parameters are partially constrained by the deformed toric ideal, without which the linear combination parameters cannot be restricted.
With the additional constraint that the total amount of enzyme is constant, we show that the concentrations and the flux are determined completely.

In FBA, steady state flux is described as the linear combination of extreme pathways.
In the previous study on FBA, its linear combination parameters can take arbitrary values and are constrained by linear inequalities (linear programming).
SNA and FBA are closely related in the sense of treating steady state flux.
Our objective is to show that the linear combination parameters cannot take arbitrary values and they are automatically constrained by deformed toric ideal constraints, after the introduction of simple mass action kinetics.

This paper is organized as follows.
In section 2, we review the basic formulation of stoichiometric network analysis.
In section 3, we apply the arguments of SNA to an example of chemical reaction network.
In section 4, we discuss deformed toric ideal constraints.
In section 5, we apply the arguments of SNA and deformed toric ideal constraint to an example of metabolic pathways.
In section 6, we give the conclusions of the analysis of the current paper and the future directions of mathematical studies of stoichiometric networks.

%%%%%%%%%
\section{Stoichiometric Network Analysis and Flux Balance Analysis}
%%%%%%%%%

The stoichiometric network analysis (SNA) starts with the chemical reaction systems which are described by

\begin{equation}
\dot{x}=S\cdot v(x;k).
\label{stoic}
\end{equation}
Here, {$x$} and {$\dot{x}$} are a concentration vector of reactant species and its time derivative, respectively, $S$ is the stoichiometric matrix, and $v(x;k)$ is the flux, where $k$ is reaction coefficients.
The stoichiometric matrix, explained below, can be determined, once we know the form of chemical equations,

\begin{equation}
a_{1j}X_1+\cdots +a_{mj}X_m \xrightarrow {k_j} b_{1j}X_1+\cdots +b_{mj}X_m,\; j=1,\ldots ,l ,
\label{chemical}
\end{equation}
where $k_j$ is the reaction coefficient for reaction $j$.
The form of flux vector $v(x;k)$ is determined by the mass action kinetics.
By the mass action kinetics, the velocity of reaction is described by
\begin{equation}
k_j[X_1]^{a_{1j}}\cdots[X_m]^{a_{mj}},
\end{equation}
where $k_j$ denotes the reaction coefficients and $[X_i]$ denotes the concentration of $X_i$.
With the matrices $A$ and $B$ of the elements $a_{ij}$ and $b_{ij}$, respectively, $S$ is defined as

\begin{equation}
S=B-A.
\end{equation}

The stoichiometric equation (\ref{stoic}) shows that the time derivative of concentrations of reactant species is represented by the product of the stoichiometric matrix and flux vector.

The steady state flux for a chemical reaction (dynamical) system is defined by the concentrations whose time derivatives vanish (${\dot x} =0$), i.e.

\begin{equation}
Sv(x;k)\equiv SJ=0,
\label{extreme}
\end{equation}

\noindent
where $J\equiv v(x;k)$.
When there is no confusion, we use $J$ hereafter.
The solution of $J$ that satisfies eq.(\ref{extreme}) forms a convex polyhedral cone.
Minimal generating vectors of the steady state flux are called `extreme currents.'
Each extreme current is the generator of the convex polyhedral cone.
A nonnegative linear combination of the extreme currents is also a steady state flux.

Following is the definition of a convex polyhedral cone~\cite{cox3}.
\begin{definition}
A convex polyhedral cone in $N_{\mathbb{R}}$ is a set of the form\\
\begin{equation}
\sigma = {\rm Cone}(S)=\left \{ \sum_{u\in S} \lambda _uu|\lambda _u \geq 0 \right \}\subseteq N_{\mathbb{R}}
\end{equation}
where $S\subseteq N_{\mathbb{R}}$ is finite.
We say that $\sigma$ is generated by $S$ and call $u$ as generators.
\end{definition}

In the field of metabolic pathway analysis, there is Flux Balance Analysis (FBA).
SNA and FBA are equivalent in the sense of treating null space generated by generators.
However, the interpretations of generators are different.
In SNA, the generators are called as extreme currents and they can take only positive value.
In FBA, the generators are called extreme pathways and they are permitted to take negative values which correspond to the exchange fluxes.
The exchange fluxes are the fluxes exchanged with the outer environment of the system.
For the detail of FBA, see refs.~\cite{palsson-book1,FBA}, for example.
For the comparison of these approaches, see ref.~\cite{palsson1}. 

%%%%%%%%%%%
\section{Example of a part of glycolysis}
%%%%%%%%%%%
%{\large \bf Example 2.1(Gatermann et. al. 2005 \cite{gatermann4})}

We consider the phosphofructokinase reaction which is a part of glycolysis~\cite{gatermann4}.
It is an extension of a reaction system proposed by Sel'kov '68~\cite{selkov}.
There are $m=3$ chemical species; $X_1$ denotes the product Fructose-1,6-biphosphate, $X_2$ denotes the reactant Fructose-6-phosphate and the extension $X_3$ stands for another intermediate in equilibrium with Fructose-6-phosphate.
The $l=7$ reaction laws are given by
\begin{eqnarray}
2X_1+X_2\xrightarrow {k_1} 3X_1 \nonumber\\
X_2 \displaystyle \mathop{\rightleftharpoons}^{k_4}_{k_5} 0 \mathop{\rightleftharpoons}^{k_3}_{k_2} X_1 \mathop{\rightleftharpoons}^{k_7}_{k_6} X_3. \nonumber
\end{eqnarray}
We arrange the left and right hand side of reaction laws (\ref{chemical}) in so-called {\it complexes} $C_j$, $j=1,\ldots ,6$ with the rate constants $k_{ij}>0$;
\begin{eqnarray}
C_1\xrightarrow {k_{21}} C_2 \nonumber\\
C_5 \displaystyle \mathop{\rightleftharpoons}^{k_{65}}_{k_{56}} C_6 \mathop{\rightleftharpoons}^{k_{46}}_{k_{64}} C_4 \mathop{\rightleftharpoons}^{k_{34}}_{k_{43}} C_3. \nonumber
\end{eqnarray}

\begin{figure}[h]
  \centering
  \includegraphics[width=0.3\textwidth]{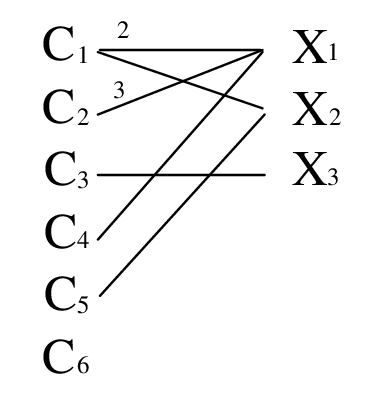}
  \caption{Stoichiometric Graph}
  \label{stoichio_graph}
\end{figure}
By the mass action kinetics, this reaction system is described by the following differential equations,

\begin{eqnarray}
\dot{x_1}&=&k_{21}x_1^2x_2+k_{46}-k_{64}x_1-k_{34}x_1+k_{43}x_3,\label{ODE_1}\\
\dot{x_2}&=&-k_{21}x_1^2x_2+k_{56}-k_{65}x_2,\label{ODE_2}\\
\dot{x_3}&=&k_{34}x_1-k_{43}x_3.\label{ODE_3}
\end{eqnarray}
Represented by the stoichiometric equation form, $S$ and $v(x;k)$ in eq.(\ref{stoic}) are given by
\begin{eqnarray}
S=
\left(
\begin {array}{ccccccc}
1&1&-1&0&0&-1&1\\
-1&0&0&1&-1&0&0\\
0&0&0&0&0&1&-1
\end {array}
\right)
\end{eqnarray}
and

\begin{eqnarray}
v(x;k)\equiv J=
\left(
\begin{array}{c}
k_{21}x_1^2x_2\\
k_{46}\\
k_{64}x_1\\
k_{56}\\
k_{65}x_2\\
k_{34}x_1\\
k_{43}x_3
\end{array}
\right).
\label{zz}
\end{eqnarray}
%
%%%%%%%%%%%%%%%%%%%
%\section{Stoichiometric Flux}
%%%%%%%%%%%%%%%%%%%
%
%\noindent
%{\large \bf Example 2.1(continued)}
The extreme currents satisfying eq.(\ref{extreme}) are given by

\begin{eqnarray}
E_1=\left (
\begin{array}{c}
0\\
1\\
1\\
0\\
0\\
0\\
0\\
\end{array}
\right )
,
E_2=\left (
\begin{array}{c}
0\\
0\\
0\\
1\\
1\\
0\\
0\\
\end{array}
\right )
,
E_3=\left (
\begin{array}{c}
0\\
0\\
0\\
0\\
0\\
1\\
1\\
\end{array}
\right )
,
E_4=\left (
\begin{array}{c}
1\\
0\\
1\\
1\\
0\\
0\\
0\\
\end{array}
\right ).
\end{eqnarray}
These are computed by setting the condition $SE_i=0$ with mathematical software Maple.
If we take the linear combination of the extreme currents,

\begin{eqnarray}
J&=&j_1E_1+j_2E_2+j_3E_3+j_4E_4\\
 &=&
\left (
\begin{array}{c}
j_4\\
j_1\\
j_1+j_4\\
j_2+j_4\\
j_2\\
j_3\\
j_3\\
\end{array}
\right ),
\label{fluxJ}
\end{eqnarray}
where, $j_l$ are nonnegative linear combination coefficients.
Thus, $J$ is a general steady state flux of eq.(\ref{extreme}).
While the coefficient $j_l$ might look to take an arbitrary nonnegative value, they have some constraints as will be shown in the next section.

%%%%%%%%%%%%%%%%%%%%
\section{Deformed Toric Ideal Constraints on Stoichiometric Network}
%%%%%%%%%%%%%%%%%%%%
The main objective of the current paper is to show that the linear combination parameters of extreme currents or extreme pathways cannot take arbitrary values.
By the introduction of mass action kinetics, they are automatically constrained and the form of flux is also constrained.
Deformed toric ideal is the main tool to derive such constraints.

By the explicit monomial vector form, the steady state flux forms a deformed toric ideal.
This can be seen by describing the elements of $J$ as monomials in the original $x$ coordinates (in affine space).
The ideal $I_{Y_L}^{def}=\{f\in \mathbb{R}[v]|f(v(x))\equiv 0\}\subseteq \mathbb{C}(k)[v]$ is called a deformed toric ideal, where $Y_L$ is the matrix whose rows are the exponent vectors of the monomials in the flux vector and its variety is called deformed toric variety.

$\mathbb{R}[v]$ is a polynomial ring, i.e. the polynomials in eq. (\ref{ODE_1})-(\ref{ODE_3}) are the elements of $\mathbb{R}[v]$.

The name `toric' results from the fact that a toric variety is invariant with respect to the induced representation of the algebraic torus~\cite{sensse}.

Following is the definition of ideal~\cite{cox1}.
\begin{definition}
A subset $I\subset k[x_1,\ldots , x_n]$ is an ideal if it satisfies:\\
(i) $0\in I$.\\
(ii) If $f,g\in I$, then $f+g\in I$.\\
(iii) If $f\in I$ and $h\in k[x_1,\ldots ,x_n]$, then $hf\in I$.\\
Here, $k$ is a field and $x_i$ are indefinite variables.
\end{definition}

\noindent
From the ref.~\cite{sturmfels}, the description of toric ideal $I\subset \mathbb{C}[x_1,\ldots, x_s]$ is\\
\begin{equation}
I_\mathcal{A}=\langle x^u-x^v | u, v \in \mathbb{N}^n, \pi(u)=\pi(v) \rangle.
\end{equation}
Here, $\mathbb{C}$ is complex number field and $\mathbb{N}^n$ is $n$-dimensional natural number space.
$\pi(u)$ is homomorphism of $u$.
This description can be written as
\begin{equation}
I_\mathcal{A}=\langle x^{u^+}-x^{u^-}: u\in {\it ker}(\pi)\rangle
\end{equation}
Here, $u^+$ and $u^-$ are non-negative support.
As we will see, the deformation by the reaction coefficients $k_{ij}$; from binomial of pure indeterminates to those with the reaction coefficients, is needed to realize the kernel of $\pi$.
The word `deformed' indicates the parameter dependence of the binomials on the reaction coefficients $k_{ij}$~\cite{sensse}.

%We will see that the form of flux is constrained by the deformed toric ideal.
%For the mathematical basics, see refs.~\cite{cox1,cox2} and~\cite{sturmfels}.

\vspace{12pt}
\noindent
{\large \bf Example of a part of glycolysis (continued)}

For the current example, the generators of the deformed toric ideal are obtained as a binomial form, and it is easy to calculate from eq.(\ref{zz});
\begin{equation}
I_J=\langle J_2-k_{46},k_{21}J_5J_3^2-k_{64}^2k_{65}J_1,k_{34}J_3-k_{64}J_6,J_4-k_{56}\rangle.
\label{idealZ}
\end{equation}
The ideal consists of the polynomial relations among the monomial coordinates of the flux vector $v(x;k)$; this is why the generators can always be taken to be binomials.

By the correspondence between the elements of vector $J$ in the representation by $J_l$ and by $j_l$, the deformed toric ideal can be described by $j_l$ coordinates.
By substituting (\ref{fluxJ}) with (\ref{idealZ}), we obtain deformed toric ideal in $j_l$ coordinates.

\begin{equation}
I_j=\langle j_1-k_{46},k_{21}j_2(j_1+j_4)^2-k_{64}^2k_{65}j_4,k_{64}j_3-k_{34}(j_1+j_4),j_2+j_4-k_{56}\rangle
\label{idealJ}
\end{equation}

\noindent
In the rest of this section, we show that $j_l$ are constrained by the deformed toric ideal.
As a result, $j_l$ are determined without introducing ad hoc constraints.
Note that in the following derivation we use only the generators of deformed toric ideal, that is, the relationships between monomials.
This has not been used in any previous studies on this example.
From the constraints that each generators of deformed toric ideal are zero, we obtain the following expressions of $j_l$,
\begin{eqnarray}
j_{1}&=&k_{46},\nonumber \\
j_{2}&=&-Z +k_{56},\nonumber \\
j_{3}&=&k_{34} ( k_{{46}}+Z)/k_{64},
\label{jl}
\\
j_{4}&=&Z.\nonumber
\end{eqnarray}

\noindent
Here, $Z$ is the solution of the following algebraic equation of degree $3$,
\begin{eqnarray}
&k_{21}Z^3+ (-k_{21}k_{56}+2k_{21}k_{46})Z^2\nonumber \\
&+ ( k_{21}k_{46}^2+k_{64}^2k_{65}-2k_{21}k_{46}k_{56}) Z-k_{21}k_{46}^2k_{56}=0.
\label{eqZ}
\end{eqnarray}
For some values of $k_{ij}$, $Z$ could be negative.
However, we consider the limited parameter region because of the non-negativity of $j_l$.
%The positivity conditions for $j_l$ give
%\begin{equation}
%0<Z<k_{56}.
%\end{equation}

\noindent
Eq.(\ref{jl}) shows that the linear coefficients $j_l$ depend on the value  of reaction rates $k_{ij}$ and $Z$.
Eq.(\ref{eqZ}) shows that $Z$ also depends on the value of $k_{ij}$, and thus $j_l$ depends on the value of $k_{ij}$, which means $j_l$ cannot be chosen arbitrarily, indifferent to the value of $k_{ij}$.
Note that these constraints are derived by the algebraic property of monomials, without introducing any ad hoc constraints, for example~\cite{loopless}.
This strong constraint seen in the current example does not hold for general reactions: the example has many generators in the deformed toric ideal enough to constrain all $j_l$ by $k_{ij}$.
In general cases, $j_l$ are only partially constrained by $k_{ij}$.

Substituting $j_1,\ldots ,j_4$ of eqs.(\ref{jl}) for eq.(\ref{fluxJ}), we obtain the stoichiometric flux $J$ under deformed toric ideal constraints,
\begin{eqnarray}
J=\left (
\begin{array}{c}
Z\\
k_{46}\\
k_{46}+Z\\
k_{56}\\
-Z +k_{56}\\
k_{34}( k_{{46}}+Z)/k_{64}\\
k_{34}( k_{{46}}+Z)/k_{64}\\
\end{array}
\right ).
\label{JJ1}
\end{eqnarray}
This means that we can determine the flux, once we know the value of reaction coefficients $k_{ij}$.% by means of experimental observation.

From eq.(\ref{zz}) and (\ref{JJ1}), we obtain the concentration as the function of reaction coefficients,
\begin{eqnarray}
x_1&=&(k_{{46}}+Z)/k_{64},\label{sol_1}\\
x_2&=&(-Z +k_{56})/k_{65},\label{sol_2}\\
x_3&=&k_{34}( k_{{46}}+Z)/k_{43}k_{64},\label{sol_3}
\end{eqnarray}
under the constraint,
\begin{equation}
Z=k_{21}(k_{{46}}+Z)(-Z +k_{56})/k_{64}k_{65}.
\end{equation}
This constraint is derived from the relation between the first ,third and fifth elements of the flux (\ref{zz}).
The concentrations of steady state are also determined completely from the value of reaction coefficients as shown above.

The concentrations obtained above are different from what are obtained by solving the equations of setting ODEs (\ref{ODE_1})-(\ref{ODE_3}) to zero.
This means the constraints used in the above derivation of concentrations (the implicit relations among the elements of flux) are stronger conditions which are constrained in solving the equations that sets ODEs to zero.

%%%%%%%%%%%%%%%%%%%
\section{Example of Feedback inhibition of pathway}
%%%%%%%%%%%%%%%%%%%

In the above sections, we considered a chemical reaction network as an example.
The same argument holds for metabolic pathways.
In this section, we consider deformed toric ideal constraints with a concrete example of metabolic pathway \cite{palsson-book2}.

%\vspace{12pt}
%\noindent
%{\large \bf Example 5.1: \\
%Feedback Inhibition of pathway, Palsson (2011) \cite{palsson-book2}}
\begin{figure}[h]
  \centering
  \includegraphics[width=1\textwidth]{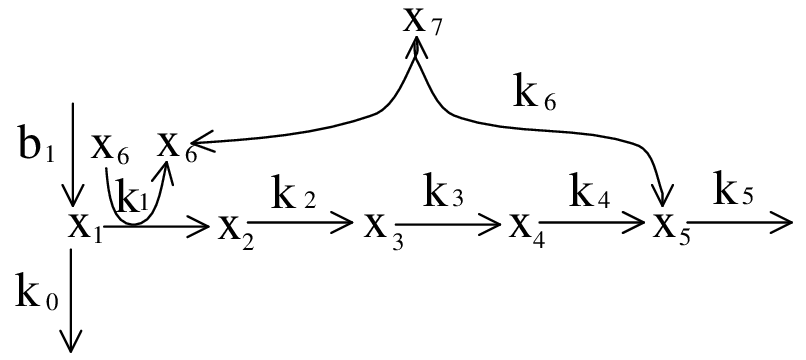}
  \caption{Feedback Inhibition of pathway}
  \label{feedback}
\end{figure}

\noindent
We use this example because it is one of the simplest realistic pathways whose monomial vector form of the flux is explicitly known.

In a biosynthetic pathway, the first reaction is often inhibited by the end product of the pathway.
Fig.\ref{feedback} illustrates a protypical feedback loop in a biosynthetic pathway.

A metabolic intermediate $x_1$ is formed and degraded as
\begin{equation}
\xrightarrow []{b_1} x_1 \xrightarrow []{k_0}.
\end{equation}
Then, if an enzyme $x_6$ is expressed, $x_1$ can be converted to $x_2$:

\begin{equation}
x_1+x_6 \xrightarrow []{k_1} x_2 +x_6
\end{equation}
which is followed by a series of reactions

\begin{equation}
x_2 \xrightarrow []{k_2} x_3 \xrightarrow []{k_3} x_4 \xrightarrow []{k_4} x_5 \xrightarrow []{k_5}
\end{equation}
to form $x_5$, the end product of the pathway.
The end product has inhibitory feedback to the enzyme $x_6$ by binding to it and converting it into an inactive form:

\begin{equation}
x_6+x_5 \displaystyle \mathop{\rightleftharpoons}_{k_{-6}}^{k_6} x_7\label{reac_rev}
\end{equation}
This system represents a simple negative feedback loop.
The differential equations that describe this feedback loop are

\begin{eqnarray}
\dot{x_1}&=&b_1-k_0x_1-k_1x_6x_1,\label{eq_x_1}\\
\dot{x_2}&=&k_1x_6x_1-k_2x_2,\label{eq_x_2}\\
\dot{x_3}&=&k_2x_2-k_3x_3,\label{eq_x_3}\\
\dot{x_4}&=&k_3x_3-k_4x_4,\label{eq_x_4}\\
\dot{x_5}&=&k_4x_4-k_5x_5-(k_6x_5x_6-k_{-6}x_7),\label{eq_x_5}\\
\dot{x_6}&=&-k_6x_5x_6+k_{-6}x_7,\label{eq_x_6}\\
\dot{x_7}&=&k_6x_5x_6-k_{-6}x_7.\label{eq_x_7}
\end{eqnarray}
In the above equations, RHS is the sum of reaction rates.
In ref.\cite{palsson-book2} (chapter 2.2), reaction rates are described mathematically using kinetic theory.
He discusses mass action kinetics as one of the fundamental concept of kinetic theory.

For the current example, the stoichiometric matrix is

\begin{equation}
S= \left(
\begin {array}{ccccccccc}
-1&0&-1&0&0&0&0&0&1\\
1&0&0&-1&0&0&0&0&0\\
0&0&0&1&-1&0&0&0&0\\
0&0&0&0&1&-1&0&0&0\\
0&-1&0&0&0&1&-1&1&0\\
0&-1&0&0&0&0&0&1&0\\
0&1&0&0&0&0&0&-1&0
\end {array}
\right)
,
\end{equation}
and the flux vector is

\begin{equation}
J= \left(
\begin {array}{c}
k_{{1}}x_{{6}}x_{{1}}\\
k_{{6}}x_{{5}}x_{{6}}\\
k_{{0}}x_{{1}}\\
k_{{2}}x_{{2}}\\
k_{{3}}x_{{3}}\\
k_{{4}}x_{{4}}\\
k_{{5}}x_{{5}}\\
k_{{-6}}x_{{7}}\\
b_{{1}}
\end {array}
\right)
.
\label{ZZ2}
\end{equation}
Notice that the third, seventh and ninth elements of flux are exchange fluxes.
These elements are exchanged from outer system.

For the metabolic pathway analysis, extreme currents correspond to extreme pathways \cite{palsson1}.%, though there is the difference in the interpretation of the exchange fluxes.
The extreme pathways computed from the stoichiometric matrix are

\begin{eqnarray}
E_1=\left(
\begin {array}{c}
0\\
1\\
1\\
0\\
0\\
0\\
0\\
1\\
1
\end {array}
\right)
,
E_2= \left(
\begin {array}{c}
1\\
0\\
-1\\
1\\
1\\
1\\
1\\
0\\
0
\end {array}
\right)
,
E_3=\left(
\begin {array}{c}
0\\
1\\
0\\
0\\
0\\
0\\
0\\
1\\
0
\end {array}
\right).\label{e_3_org}
\end{eqnarray}
By taking a linear combination of the extreme pathways, the metabolic flux is obtained;

\begin{eqnarray}
J&=&j_1E_1+j_2E_2+j_3E_3\nonumber\\
&=& \left(
\begin {array}{c}
j_2\\
j_1+j_3\\
j_1-j_2\\
j_2\\
j_2\\
j_2\\
j_2\\
j_1+j_3\\
j_1
\end {array}
\right)
.
\label{JJ2}
\end{eqnarray}
%In ref.~\cite{palsson-book2}, deformed toric ideal is not discussed.
Here, we derive the deformed toric ideal of this pathway.
From the monomial vector representation of $J$, the deformed toric ideal is given by

\begin{equation}
I_J=\langle  J_1J_7k_6k_0-J_2J_3k_1k_5\rangle.
\end{equation}
From the corresponding representation of flux, the deformed toric ideal represented by $j_l$ is given by

\begin{equation}
I_j=\langle j_2^2 k_6k_0-(j_1+j_3)(j_1-j_2) k_1k_5\rangle.
\end{equation}

There is only one deformed toric ideal constraint, which is obtained by equating the generator to zero.
%This is a relation between the extreme pathways.
The parameter region is partially constrained by the reaction coefficients.
%In metabolic pathways, the flux is important.
The flux is partially constrained by experimental observations, furthermore automatically constrained by the deformed toric ideal constraints.

In the above example, there is only one deformed toric ideal constraint.
Comparing with the example of chemical reaction network, the number of constraints is small.
This is caused by the small number of the generators of the deformed toric ideal.
The smallness originates in the limited number of species which appear as the same form in the chemical equations.

%The points of $j_1=j_2$, $j_2=0$ or $j_1+j_3=0$ are critical points between the thermodynamically feasible and infeasible region in the sense of refs.~\cite{qian1,qian2},
%where thermodynamic feasibility is described as the orthogornality to the cycles of the oriented matroid associated with stoichiometric matrix $S$.
%At the above points, the elements of flux is zero.
%For the detail, see ref.~\cite{sato}.
In the above derivation, the solution space of the equations, which equate eqs.(\ref{eq_x_1})-(\ref{eq_x_7}) to zero, considers the solution in 7-dimensional space ($x$-coordinates).
In $j$-coordinates, the solution space is reduced to 3-dimension and the only one deformed toric ideal constraint gives the 2-dimensional hypersurface in this space.
Thus, we are considering the hypersurface in the reduced dimensional space and we will consider the solution on this hypersurface, below.

In the rest, we will show that, in addition to the steady state equations or mass balance, which were already known in ref.~\cite{palsson-book2}, the above deformed toric ideal constraint provides the complete solution to the steady state of this system.  First, as in \cite{palsson-book2}, the steady state equations are given by 

\begin{eqnarray}
0&=&b_1-k_0x_1-k_1x_6x_1,\label{eq_s_x_1}\\
0&=&k_1x_6x_1-k_2x_2,\\
0&=&k_2x_2-k_3x_3,\\
0&=&k_3x_3-k_4x_4,\\
0&=&k_4x_4-k_5x_5-(k_6x_5x_6-k_{-6}x_7),\\
0&=&-k_6x_5x_6+k_{-6}x_7,\\
0&=&k_6x_5x_6-k_{-6}x_7.\label{eq_s_x_7}
\end{eqnarray}
Here, we introduce the mass balance in the sum of the enzyme, $x_6$ and $x_7$:
%Next, the sum of $x_6$ and $x_7$ gives us the mass balance in the enzyme:

\begin{equation}
x_6+x_7=e_t,
\label{total_enzyme}
\end{equation}
where $e_t$ is the total amount of enzyme (constant).

It is known that these equations (\ref{eq_s_x_1})-(\ref{total_enzyme}) can be combined to give a quadratic equation,

\begin{equation}
y^2+ay-b=0,
\label{combined}
\end{equation}
where

\begin{equation}
y=k_2x_2, a=k_5k_{-6}(1+k_1 e_t/k_0)/k_6, b=k_5k_{-6}k_1 e_tb_1/(k_6k_0),
\end{equation}
that has one positive root in $y$, because $(-a+\sqrt{a^2+4b})/2>0$.
Note that from $y=k_2x_2=J_4=j_2$ the variable 
$j_2$ is given by a positive root of eq.(\ref{combined}).
Eq.(\ref{combined}) holds because of the unique property of this dynamical system.
This does not hold for general system.

In ref.~\cite{palsson-book2}, these constraints are not studied with $j_l$ coordinates.
With the representation of flux by the extreme pathways, eq.(\ref{total_enzyme}) can be represented by $j_l$ coordinates, eq.(\ref{total_enzyme}) gives
\begin{equation}
(j_1+j_3)k_5/j_2k_6+(j_1+j_3)/k_{-6}=e_t.
\end{equation}
With these conditions, the region taken by the parameters $j_l$ is further limited, combined with the deformed toric ideal constraint.% and nonnegativity conditions on $j_l$.
\begin{eqnarray}
j_1&=&j_2 ( k_6k_0+k_1k_5 )/k_1k_5,\label{j_1_res}\\
j_3&=&\frac {j_2 \{k_5k_{-6}(e_tk_1k_6-k_1k_5-k_6k_0)-j_2(k_1k_5k_6+k_6^2k_0) \} }{k_1k_5 ( k_5k_{-6}+j_2k_6 ) }.%\geq 0.
\label{j_3_res}
\end{eqnarray}
Since from eq.(\ref{combined}) $j_2$ is determined uniquely, eqs.(\ref{j_1_res}), (\ref{j_3_res}) shows $j_1, j_3$ are also unique.
%In addition, when the RHS of eq.(\ref{j_3_res}) is nonnegative, a unique solution for $(j_1,j_2,j_3)$ is determined, and when the RHS is negative, there is no steady state flux.

From eq. (\ref{ZZ2}) and (\ref{JJ2}), the concentrations are
\begin{eqnarray}
x_1&=&(j_1-j_2)/k_0,\\
x_2&=&j_2/k_2,\\
x_3&=&j_2/k_3,\\
x_4&=&j_2/k_4,\\
x_5&=&j_2/k_5,\\
x_6&=&k_5(j_1+j_3)/k_6j_2=k_0j_2/k_1(j_1-j_2),
\label{x_6}\\
x_7&=&(j_1+j_3)/k_{-6},\\
b_1&=&j_1,
\end{eqnarray}
here, RHS of eq.(\ref{x_6}) is the constraint which is derived from the relation between the first, second, third and seventh elements of the flux (\ref{ZZ2}).

The concentrations are determined uniquely by solving the eqs. (\ref{j_1_res}), (\ref{j_3_res}) and RHS of (\ref{x_6}).
We obtain,
\begin{eqnarray}
j_1&=&\frac {k_{-6} (-k_5+e_t k_6)( k_6k_0+k_1k_5)}{k_1k_5k_6},\\
j_2&=&\frac {k_{-6} (-k_{5}+e_{t}k_{6})}{k_{6}},\\
j_3&=&-\frac {k_{-6}k_0 (-k_{{5}}+e_{t}k_{6})}{k_1k_5}.\label{j_3_neg}
\end{eqnarray}
Therefore, the concentrations are determined as
\begin{eqnarray}
x_1&=&{\frac {k_{{-6}} \left( -k_{{5}}+e_{{t}}k_{{6}} \right) }{k_{{1
}}k_{{5}}}},\\
x_2&=&\frac {k_{-6} (-k_{5}+e_{t}k_{6})}{k_{6}k_2},\\
x_3&=&\frac {k_{-6} (-k_{5}+e_{t}k_{6})}{k_{6}k_3},\\
x_4&=&\frac {k_{-6} (-k_{5}+e_{t}k_{6})}{k_{6}k_4},\\
x_5&=&\frac {k_{-6} (-k_{5}+e_{t}k_{6})}{k_{6}k_5},\\
x_6&=&k_5/k_6,
\label{x_6_1}\\
x_7&=&{\frac {-k_{{5}}+e_{{t}}k_{{6}} }{k_{{6}}}},
\label{x_7_1}\\
b_1&=&\frac {k_{-6} (-k_5+e_t k_6)( k_6k_0+k_1k_5)}{k_1k_5k_6}.
\end{eqnarray}
These hold for $e_t k_6>k_5$ from positivity conditions of concentrations.
Unless, these give the negative concentrations.
This contradicts the positivity of concentrations.
The concentrations of $x_1$ to $x_5$ are the same.
These correspond to the series along the horizontal arrows of figure 2.

In eq.(\ref{j_3_neg}), $j_3$ is negative.
Note that the reaction corresponding to $E_3$ is reversible, because $E_3$ in eq.(\ref{e_3_org}) has the second and eighth elements and these correspond to $x_5, x_6$ and $x_7$ in eq.(\ref{ZZ2}).
From eq.(\ref{reac_rev}), the reaction including these metabolites is reversible.
Therefore $j_3$ can take negative value.

We can also determine the flux explicitly as
\begin{eqnarray}
J=
\left(
\begin {array}{c}
k_{-6} (-k_5+e_t k_6)/k_6\\
k_{-6} (-k_5+e_t k_6)/k_6\\
k_{-6}k_0( -k_5+e_t k_6)/k_1k_5\\
k_{-6} (-k_5+e_t k_6)/k_6\\
k_{-6} (-k_5+e_t k_6)/k_6\\
k_{-6} (-k_5+e_t k_6)/k_6\\
k_{-6} (-k_5+e_t k_6)/k_6\\
k_{-6} (-k_5+e_t k_6)/k_6\\
k_{-6} (-k_5+e_t k_6)( k_6k_0+k_1k_5)/k_1k_5k_6
\end {array}
\right)
.
\end{eqnarray}
This is the flux at steady state.
Although, in eq.(\ref{j_3_neg}), $j_3$ is negative, $j_3$ appears as the form of $j_1+j_3$ in $J$.
Therefore all elements of $J$ are non-negative.

Except for the third and ninth elements, the other elements are the same.
The third and ninth elements correspond to the exchange fluxes.
In the steady state, the other fluxes take the same value.

Although the seventh element is one of the exchange fluxes, the flux takes the same value as those of internal fluxes.
This is caused by the equilibrium between the effects of internal fluxes and of exchange fluxes.
The effects of internal fluxes may be stronger than those of exchange fluxes.
Therefore the seventh element takes the same value as the internal fluxes.

%%%%%%%%%%%%%%%%%%%
\section{Conclusions}
%%%%%%%%%%%%%%%%%%%

In this paper, we considered deformed toric ideal constraints on stoichiometric networks, treating the chemical reaction networks and metabolic pathways in a unified way.
This paper is the first that pointed out that the deformed toric ideal constrains the linear combination parameters of flux.
We have seen that the steady state fluxes are constrained by the deformed toric ideals and obtained the explicit form of concentrations.

In general, for metabolic pathways, reactants and products are so diverse that the elements of flux may not appear as one variable.
The number of generators for the deformed toric ideal is thus small and the effect of constraints may be limited.
By considering sub-networks of large scale metabolic pathways, however, they should be more influential.
It will be interesting to consider sub-networks with many exchange fluxes, because there appear some elements of fluxes as one variable.

In refs.~\cite{qian1,qian2}, thermodynamic constraints are considered explicitly, in terms of the non-equilibrium thermodynamic systems.
Such constraints are not discussed in the current paper, but the study with thermodynamic constraints, added to the deformed toric ideal constraints, will be interesting.
By considering these constraints, flux is constrained in another form.

On the other hand, one can calculate deformed toric ideal from Gr\"{o}bner basis in general~\cite{sensse}.
Therefore, our analysis with deformed toric ideal is applicable even for general large scale pathways.

As another topic, which is pointed out for the example of chemical reaction network, flux forms a convex polyhedral cone.
The algebraic geometrical study, using commutative algebra and combinatorics, will also be interesting.
%This is within our future work.
\vspace{.1in}

\begin{center}
    {\Large {\bf Acknowledgements} }
\end{center}
\vspace{.1in}
We greatly thank to S. Mano and C. Miura for the fruitful discussions and careful reading of the first version of the manuscript.
We also thank to M. Arita, Y. Hasegawa and A. Mochizuki for discussions from the biological view point and from the view point of dynamical systems.
MS has been supported by JST-NSF grant, SICORP.
KF has been supported in part by JSPS KAKENHI (B) 22300098.

\pagebreak

\end{document}